\documentstyle[prb,aps,epsf]{revtex}
\begin{document}
\draft

\twocolumn[
\hsize\textwidth\columnwidth\hsize\csname@twocolumnfalse\endcsname\title
{{\it Ab initio} calculations for bromine adlayers on the Ag(100)
            and Au(100) surfaces: \\
            the $c(2\times2)$ structure}

\author {Sanwu~Wang$^{1}$ and Per~Arne~Rikvold$^{1,2}$}

\address{$^1$School of Computational Science and Information Technology,\\
             and Center for Materials Research and Technology,\\
             Florida State University, Tallahassee, Florida 32306-4120\\
         $^2$Department of Physics, Florida State University, Tallahassee,
             Florida 32306-4350}

\maketitle

\begin{abstract}

{\it Ab initio} total-energy density-functional methods with supercell
models have been employed to calculate the $c(2\times2)$ structure of the
Br-adsorbed Ag(100) and Au(100) surfaces. The atomic geometries of the
surfaces and the preferred bonding sites of the bromine have been
determined. The bonding character of bromine with the substrates has also
been studied by analyzing the electronic density of states and the charge
transfer. The calculations show that while the four-fold hollow-site
configuration is more stable than the two-fold bridge-site topology on the
Ag(100) surface, bromine prefers the bridge site on the Au(100) surface.  
The one-fold on-top configuration is the least stable configuration on
both surfaces. It is also observed that the second layer of the Ag
substrate undergoes a small buckling as a consequence of the adsorption of
Br. Our results provide a theoretical explanation for  the experimental
observations that the adsorption of bromine on the Ag(100) and Au(100)
surfaces results in different bonding configurations.

\end{abstract}

]

\section{Introduction}

Anion adsorption on metals can strongly modify surface morphology and  
electronic structure and chemical reactivity. It is therefore of great
scientific and technological importance. In particular, the
halide-adsorbed noble-metal systems play a significant role in
electrochemistry. From the fundamental point of view, halide-adsorbed
noble-metal surfaces are important model systems for adsorption on metal  
surfaces with formation of ordered two-dimensional adsorbate structures.  
It is thus not surprising that the adsorption of halides on noble metals  
has been extensively investigated.

The systems we selected to study are the Br-chemisorbed Ag(100) and
Au(100) surfaces. The adsorption of bromine on the Ag(100) and Au(100)  
surfaces both in vacuum and in solution have been widely studied by   
experiments
\cite{Kleinherbers,Ocko1,Hanewinkel,Endo2,Wandlowskinew,Bertel,Netzer,%
Ocko3,Wandlowski,Pajkossy,Ocko2,Cuesta} and by classical simulations.  
\cite{Kope1,Kope2,Mitchell1,Mitchell2} Experimentally, bromine has been
found to form different bonding structures on the Ag(100) and Au(100)    
surfaces. While bromine chemisorbed on the Ag(100) surface occupies the  
four-fold hollow site (hereafter referred to as $H_4$), the most stable
chemisorption structure on Au(100) is the configuration with bromine
at the two-fold bridge site (hereafter referred to as $B_2$). These      
different chemisorption structures have been verified by various
experimental measurements.\cite{Kleinherbers,Ocko1,Hanewinkel,Endo2,%
Bertel,Netzer,Ocko3,Wandlowski,Pajkossy,Ocko2,Cuesta} However, theoretical
studies have not yet reproduced these different adsorption behaviors.

Kleinherbers {\it et al.} performed angle-resolved photoemission,
low-energy electron diffraction (LEED), and X-ray photoemission
measurements for the interaction of halides with Ag surfaces.
\cite{Kleinherbers} They found that the adsorption of Cl, Br, and I on the
Ag(100) surface in vacuum all resulted in the formation of a $c(2\times2)$
overlayer with the adsorbates in the $H_4$ sites. Using {\it in situ}  
surface X-ray scattering, Ocko {\it et al.} studied the adsorption
of bromide on an Ag(100) electrode. They observed a disordered phase at
lower coverages and an ordered $c(2\times2)$ phase at a coverage of half a
monolayer. \cite{Ocko1} The Br was determined to bond at the $H_4$ site in
both the $c(2\times2)$ and disordered phases. The atomic geometry of the  
Br/Ag(100)-$c(2\times2)$ surface was further investigated by Endo {\it et 
al.} with the {\it in situ} X-ray absorption fine structure (XAFS) method.
\cite{Endo2} The $H_4$ site was confirmed to be the bonding site of
bromine. The disordered phase of the Br/Ag(100) surface at lower
bromine coverages was also recently further investigated experimentally.
\cite{Hanewinkel} In that study, it was suggested that while most of the
bromide ions occupy the $H_4$ sites, there are additional bromide ions
adsorbed slightly off the $H_4$ sites.

The LEED data reported by Bertel {\it et al.} \cite{Bertel,Netzer} have
shown that the chemisorption of Br on the Au(100) surface in vacuum
results in the rearrangement of the top-layer Au atoms of the original   
clean reconstructed Au(100)-$(5\times20)$ surface and the formation of an
unreconstructed $(1\times1)$ substrate structure. Several ordered
structures of the Br adlayer, including $c(2\times2)$,
$(\sqrt{2}\times4\sqrt{2})R45^\circ$, and $c(4\times2)R45^\circ$, were   
obtained after bromine exposure on Au(100) surfaces, with the former two
structures being metastable. It was concluded from the experimental data
that Br adsorbed at the $B_2$ site on the Au(100) surface in all the
observed phases. This is in contrast to the case of the Br-adsorbed  
Ag(100) surface. Under electrochemical {\it in situ} conditions, surface
X-ray scattering and scanning tunneling microscopy (STM) experiments
showed that bromide adsorbed on the unreconstructed Au(100)-$(1\times1)$
surface forms a commensurate $c(\sqrt{2}\times2\sqrt{2})R45^\circ$
structure and an incommensurate $c(\sqrt{2}\times2p)R45^\circ$ ($p \leq   
2\sqrt{2}$, depending on the applied potential).
\cite{Ocko3,Wandlowski,Ocko2,Cuesta} In this case, too, the bromide ions
were determined to reside at the $B_2$ sites.

In contrast to the considerable progress of the experimental measurements,
theoretical studies employing {\it ab initio} methods to these systems are
still at an early stage. Several groups have performed {\it ab initio}
Hartree-Fock (HF) and density-functional-theory (DFT) calculations for the
Br/Ag(100) and Br/Au(100) interfaces using cluster models.
\cite{Gomes,Ignaczak3,Pacchioni,Illas} While these investigations have
provided useful information about the  interaction between Br and the   
surfaces, as we discuss below, many of the results are not yet
sufficiently accurate. For example, the preferred bonding site of Br on
the Au(100) surface was incorrectly predicted by {\it ab initio} DFT
cluster calculations, which showed that Br would prefer to bond at the
$H_4$ site on both the Ag(100) and Au(100) surfaces.\cite{Ignaczak3}
Similarly, {\it ab initio\/} HF studies with small clusters predicted the
bridge site as the preferred adsorption site for Br on Ag(100).
\cite{Pacchioni} Given that these calculations with small clusters cannot
reproduce such a fundamental property as the binding site, all other
results ({\it e.g.}, energy barriers) obtained from such calculations for
both the Br/Ag(100) and Br/Au(100) systems are questionable.

Here we present results of total-energy DFT calculations in which we used
supercell models for the Ag(100) and Au(100) surfaces. The detailed atomic
structures and electronic properties of the chemisorbed surfaces and the
preferred bonding site of the adsorbate have been determined. Our   
theoretical approach has reproduced the different behavior of Br on the   
Ag(100) and Au(100) surfaces. The most stable  adsorption sites for Br
chemisorbed on the Ag(100) and Au(100) surfaces are determined by our  
calculations to be the $H_4$ and the $B_2$ sites, respectively. Our
results are in excellent agreement with the experimental data. The
obtained results for the electronic properties also enable us to analyze
the nature of the bonding between Br and the substrates and understand the
different adsorption behavior of Br on the Ag(100) and Au(100) surfaces.  

The remainder of this paper is organized as follows. In Sec.~II we outline
in detail the computational method and the supercell models that we used. 
In Sec.~III we present and discuss the results for bulk, clean surfaces
(Sec.~III~A), and adsorbed surfaces (Sec.~III~B). The adsorption
geometries and atomic relaxations are discussed in Sec.~III~B1, and the 
electronic properties and bonding character in Sec.~III~B2. We also give
comparisons of our results with previous calculations and experimental 
data. Finally, in Sec.~IV, we summarize the main results of our
calculations.

\section{Method and model}

On the Ag(100) and Au(100) surfaces, there are three different symmetric
adsorption sites, known as $H_4$, $B_2$, and $T_1$ (on-top) sites. These
three sites are shown in Fig.~1. We have studied a $c(2\times2)$ structure
in which Br 

\vspace{.5cm}
\begin{figure}[b]
\epsfbox{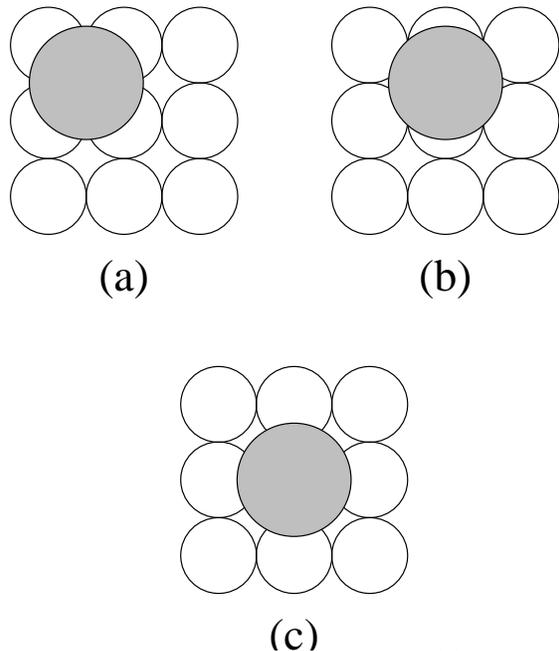}
\caption{Schematics of an adatom at (a) the four-fold hollow ($H_4$) site,
(b) the two-fold bridge ($B_2$) site, and (c) the on-top ($T_1$) site on
the unreconstructed Ag(100) and Au(100) surfaces.}
\label{autonum}
\end{figure}
\vspace{.5cm}

\parindent=0mm
forms an adlayer with a coverage of $1\over2$ monolayer on the
surface for each of the three bonding configurations.

\parindent=4mm

The metal surface is modeled by repeated slabs with five, seven, and nine
metal layers separated by a vacuum region equivalent to five or seven
metal layers. Each metal layer in the supercell contains two metal atoms.
Br is adsorbed symmetrically on both sides of the slab. All the metal
atoms were initially located at their bulk positions, with the equilibrium
lattice constant of the bulk determined by our calculations.

The calculations were performed within density-functional theory, using
the pseudopotential (PP) method and a plane-wave basis set. The results
reported in this paper were obtained using the Vienna {\it ab-initio}
simulation package (VASP). \cite{Kresse1,Kresse2,Kresse3} The 
exchange-correlation effects were treated with the generalized
gradient-corrected exchange-correlation functionals (GGA) given by Perdew
and Wang. \cite{Perdew1,Perdew2} We adopted the scalar-relativistic
Vanderbilt ultrasoft pseudopotentials supplied by Kresse and Hafner
\cite{Vanderbilt,Kresse4}. A plane-wave energy cutoff of 20 Ry and 56
special {\bf k} points in the irreducible part of the two-dimensional
Brillouin zone of the $c(2\times2)$ surface were used for calculating both
the Br/Ag(100) and Br/Au(100) surfaces. Optimization of the atomic
structure was performed for each supercell {\it via} a conjugate-gradient
technique using the total energy and the Hellmann-Feynman forces on the
atoms. \cite{Payne} All the structures were fully relaxed until the change  
in total energy was smaller than 1 meV between two ionic steps. The
convergence of the total energies was checked with different values of the
plane-wave cutoff and different numbers of special {\bf k} points. A
series of test calculations with different slab thicknesses (from five to
nine metal layers) and vacuum-gap widths (equivalent to five and seven
metal layers) were also carried out to check convergence. The calculations
on which this paper is based represent approximately 300 CPU hours on an
IBM SP2 computer.

\section{Results and discussion}

\subsection{Bulk and clean surface}

We first present the calculated properties for bulk silver and bulk gold,
and the relaxed but unreconstructed clean Ag(100) and Au(100) surfaces.  

Calculations for bulk Ag and Au were conducted with 408 special {\bf k}  
points and cutoff energies ranging from 20 Ry to 40 Ry. The total energy
convergence with respect to the cutoff energy was shown to be within a few
tenths of 1 meV. We obtained lattice constants of 4.17 {\AA} and 4.18
{\AA} for bulk Ag and Au, about 2.0\% and 2.5\% larger than the
corresponding experimental values \cite{CRC} at room temperature,
respectively. Previous total-energy DFT calculations at the GGA level
found lattice constants between 4.13 {\AA} and 4.19 {\AA} for bulk Ag,
\cite{Khein,Yu1,Ratsch,Gravil,Gravi2,Eichler,Asato,Vitos} and between 4.19
{\AA} [Ref.~\onlinecite{Yu2}] and 4.20 {\AA} [Ref.~\onlinecite{Vitos}] for
bulk Au. Our results are in good agreement with these calculations.

The properties of the clean Ag(100) and Au(100) surfaces were calculated
using supercells containing a 7-layer metal slab and a vacuum gap
equivalent to 7 bulk-metal layers. A $1\times1$ surface cell was used with
66 special {\bf k} points in the surface Brillouin zone. The kinetic
energy cutoff for the calculations was 20 Ry. All the layers except for
the central one were relaxed. Surface reconstruction was not considered.

The surface energies for the Ag(100) and Au(100) surfaces obtained from
our calculations are 0.43 eV/atom and 0.47 eV/atom, respectively. Both 
results are in good agreement with recent pseudopotential GGA calculations
by Yu and Scheffler, who reported the corresponding values of 0.48 and
0.45 eV/atom.\cite{Yu1,Yu2} A recent calculation with
linear-muffin-tin-orbital (LMTO)-GGA methods, however, obtained much
larger values of 0.65 and 0.90 eV/atom for the unrelaxed Ag(100) and
Au(100) surfaces, respectively. \cite{Vitos} The reason for the large
discrepancy from the other GGA results is not clear. Pseudopotential 
local-density approximation (LDA), \cite{Yu1,Yu2} LMTO-LDA,
\cite{Methfessel,Boisvert,Fiorentini} and linearized augmented-plane-wave
LDA \cite{Weinert,Erschbaumer,Eibler} calculations have provided values of
0.59--0.7 eV/atom, and 0.69--0.72 eV/atom for the surface energies of the 
Ag(100) and Au(100) surfaces, respectively. These DFT-LDA values for the 
surface energy are generally larger than those calculated with the DFT-GGA
calculations reported in this paper and Refs. \onlinecite{Yu1} and
\onlinecite{Yu2}. This is consistent with the previous observation
\cite{Perdew1} that LDA surface energies are normally larger than the
corresponding GGA values due to the different treatment of the
exchange-correlation functional. The calculated values for the surface
energies are thus seen to be quite sensitive to the computational method
and the form of the exchange-correlation functional.

Table~I shows the results of the surface relaxation. While no significant
structural relaxation is found for either surface, the Ag(100) surface
shows a slightly more relaxed geometry than the Au(100) surface. Both
surfaces show an inward relaxation of the top layer and slight outward
relaxation of the second and third layers. LEED measurements \cite{Li}
showed insignificant relaxation of the Ag(100) surface with ${\Delta
d_{12}}/d_0 = 0 \pm 1.5 \%$ and ${\Delta d_{23}}/d_0 = 0 \pm 1.5 \%$,
where ${\Delta d_{12}}$ and ${\Delta d_{23}}$ are the changes in spacing
between the top and the second layer and between the second  and the third
layer, and $d_0$ is the bulk interlayer distance. Our results are thus in
good agreement with the experimental data, and basically consistent with
other {\it ab initio} calculations, the results of which are also listed
in Table~I for comparison.

\subsection{Br-adsorbed Ag(100) and Au(100) surfaces}

\subsubsection{Relaxations and energetics}

The results of our calculations for the Br/Ag(100)-$c(2\times2)$ and the
Br/Au(100)-$c(2\times2)$ surfaces are shown in Tables II and III. If not
otherwise indicated, the results reported in this section (and in Tables
II and III) were calculated with a cutoff energy of 20 Ry, 56 special {\bf
k} points, and supercells containing a 9-layer metal slab 
with a vacuum
region equivalent to seven metal layers.

\vspace{0.5cm}
\begin{table}
\caption{Relaxation of the clean Ag(100) and Au(100) surfaces. $\Delta
d_{ij}$ is the change of the interlayer distance, and $d_{0}$ is the
corresponding distance in the bulk.}
\begin{tabular}{lcccc}
&${\Delta d_{12}}/d_0$ (\%)&${\Delta d_{23}}/d_0$ (\%)&${\Delta
d_{34}}/d_0$ (\%)\\
\tableline
Ag(100)&&&\\
This work&$-$1.8&0.7&0.2\\
PP-GGA\tablenotemark[1]&$-$1.4&&\\
PP-LDA\tablenotemark[1]&$-$2.2&0.4&\\
PP-LDA\tablenotemark[2]&$-$1.3&1.0&0.8\\
LMTO-LDA\tablenotemark[3]&$-$1.9&&\\
Experiment\tablenotemark[4]&$0\pm1.5$&$0\pm1.5$&\\
\tableline
Au(100)&&&\\
This work&$-$1.3&0.3&0.2\\
PP-LDA\tablenotemark[5]&$-$1.2&0.4&\\
LMTO-LDA\tablenotemark[6]&$-$1.0&&\\
\end{tabular}
\tablenotetext[1]{Ref. \onlinecite{Yu1}.}
\tablenotetext[2]{Ref. \onlinecite{Bohnen}.}
\tablenotetext[3]{Refs. \onlinecite{Methfessel,Boisvert,Fiorentini}.}
\tablenotetext[4]{Ref. \onlinecite{Li}.}
\tablenotetext[5]{Ref. \onlinecite{Yu2}.}
\tablenotetext[6]{Refs. \onlinecite{Boisvert,Fiorentini}.}
\end{table}

In-plane relaxations of the
top-layer metal atoms were found to result in changes of the distance
between Br and its nearest-neighbor metal atoms and of the total-energy
difference between two different configurations within only 0.01 {\AA} and
a few meV, respectively. The effects of in-plane relaxations are thus
negligible, and such relaxations were not considered in the calculations
for the Br-adsorbed surfaces.

Table~II shows total-energy differences between different bonding
configurations of both surfaces. Each structure was optimized. We found   
that while the total energy of the $H_4$ configuration is lower by 213 meV
than the $B_2$ configuration for the Br/Ag(100) surface, it is higher by
58 meV for the Br/Au(100) surface. The $T_1$ configuration for both
surfaces is found to be higher in total energy than both the corresponding
$H_4$ and $B_2$ configurations. Thus, we conclude that while Br adsorbed  
on the Ag(100) surface prefers the $H_4$ site, it is adsorbed at the $B_2$
site on the Au(100) surface. This conclusion is in agreement with
experimental observations. \cite{Kleinherbers,Ocko1,Hanewinkel,Endo2,%
Bertel,Netzer,Ocko3,Wandlowski,Pajkossy,Ocko2,Cuesta}

It is interesting to note that the magnitude of the total-energy
difference between the $H_4$ and $B_2$ structures for Br/Ag(100) is 
significantly larger than the corresponding value for the Br/Au(100)
surface. This suggests that diffusion of Br on the Au(100) surface may
occur much more easily than on the Ag(100) surface since the total-energy
difference between the most stable configuration (the global minimum) and
the less favorable configuration (probably a saddle point) is directly
relevant to adsorbate diffusion.

Previous theoretical studies employing cluster models also determined the
preferred bonding sites of Br on the Ag(100) and Au(100) surfaces, as
mentioned in Sec.~I. {\it Ab initio} HF calculations showed that Br would 
prefer to bond at the $B_2$ site on the Ag(100) surface (by 370 meV/adatom
over the $H_4$ site, and by 570 meV/adatom over the $T_1$ site).
\cite{Pacchioni} This is inconsistent with both our DFT-supercell
calculations and the experimental data.
\cite{Kleinherbers,Ocko1,Hanewinkel,Endo2} DFT cluster calculations  
predicted that the binding energy of Br at the $H_4$ site on both the
Ag(100) and Au(100) surfaces was larger than at the $B_2$ and $T_1$ sites
by 120 meV for Ag(100) [89 meV for Au(100)] 
and

\vspace{0.5cm}
\begin{table}
\narrowtext
\caption{Total energy differences (in eV per unit cell) between different
configurations of the Br/Ag(100)-$c(2\times2)$ and
Br/Au(100)-$c(2\times2)$ surfaces, obtained from calculations with
supercells containing a 9-layer slab and a 7-layer vacuum region, a cutoff
energy of 20 Ry, and 56 special {\bf k}-points.}
\begin{tabular}{lcccc}
$E_{H_4}-E_{B_2}$ (Br/Ag(100))&&&$-$0.213&\\
$E_{H_4}-E_{T_1}$ (Br/Ag(100))&&&$-$0.557&\\
$E_{B_2}-E_{T_1}$ (Br/Ag(100))&&&$-$0.344&\\
\tableline
$E_{H_4}-E_{B_2}$ (Br/Au(100))&&&$+$0.058&\\
$E_{H_4}-E_{T_1}$ (Br/Au(100))&&&$-$0.244&\\
$E_{B_2}-E_{T_1}$ (Br/Au(100))&&&$-$0.302&\\
\end{tabular}
\end{table}

\parindent=0mm

202 meV for Ag(100) [202
prefers to bond
at the $H_4$ site on the Au(100) surface is in
disagreement with experimental measurements,
\cite{Bertel,Netzer,Ocko3,Wandlowski,Pajkossy,Ocko2,Cuesta}
as well as with our results. We believe that the main problem is that 
these previous calculations were limited to small clusters, containing
only up to 13 metal atoms. It is well known that a small metal cluster has
a very different electronic structure than an extended metal surface,  
yielding very significant differences in adsorbate binding energies and
reaction pathways.\cite{Whitten,Panas,Whetten,Gensic} Large clusters or 
extended surface models (e.g., supercell models) are therefore needed to
simulate real metal surfaces accurately.

\parindent=4mm

The structural parameters of the optimized geometries for the $H_4$,
$B_2$, and $T_1$ configurations of the Br/Ag(100) and Br/Au(100) surfaces
are presented in Table~III. The vertical distances ($d_z$) between the Br
centers and the plane of the centers of the top-layer atoms were
calculated to be 1.91 {\AA} on the Ag(100) surface and 2.01 {\AA} on the
Au(100) surface for the $H_4$ structure, 2.16 {\AA} on Ag(100) and 2.18
{\AA} on Au(100) for the $B_2$ configuration, and 2.48 {\AA} on Ag(100)
and 2.46 {\AA} on Au(100) for the $T_1$ structure. While the values of
$d_z$ for the $B_2$ and $T_1$ configurations of the Br/Ag(100) surface are
very close to the corresponding values for the Br/Au(100) surface, the
distance between Br and the surface in the $H_4$ configuration is observed
to be significantly longer (by 0.1 {\AA}) on Br/Au(100) than on
Br/Ag(100). Accurate {\it in-situ} XAFS measurements for Br/Ag(100) in   
NaBr solution by Endo {\it et al.} showed that the bond length between Br
and its four nearest-neighbor Ag atoms in the $H_4$ configuration is   
$2.82\pm0.05$ {\AA}, and the distance between the Br and the surface is  
$1.94\pm0.07$ {\AA}. \cite{Endo2} Our results (2.82 {\AA} and 1.91 {\AA},
respectively) are thus in excellent agreement with the experimental data,
provided that the solution has only a minor influence on the bond lengths
between the adsorbate and the surface.

The bond lengths between Br and Ag and Au clusters of varying size have
been obtained with both HF and DFT calculations. Illas {\it et al.}, using
the HF method with a cluster of 5 Ag atoms simulating the $H_4$
configuration of the Ag(100) surface, obtained a value of 3.43 {\AA} for
the length of the Br-Ag bond.\cite{Illas} Paccioni, also using the HF
method with slightly larger clusters, found that the bond lengths between
a Br ion and the surface Ag atom were 3.24 {\AA}, 2.97 {\AA}, and 2.94
{\AA} in clusters of Br$^-$-Ag$_{13}$ (modeling the $H_4$ structure),
Br$^-$-Ag$_{8}$ (simulating the $B_2$ geometry), and Br$^-$-Ag$_{13}$   
(representing the $T_1$ configuration), respectively. Ignaczak and Gomes 
performed DFT calculations with clusters containing a Br ion and 12 metal
atoms and determined the bond lengths to be 3.2 {\AA}, 3.0 {\AA}, and 2.9
{\AA} for the $H_4$, $B_2$, and $T_1$ configurations of Br$^-$-Ag$_{12}$
and Br$^-$-Au$_{12}$ clusters, respectively. All of these values are much
larger that those obtained from our supercell calculations and the XAFS
measurements, suggesting that small clusters do not represent the metal
surfaces properly.

The $B_2$ configuration of the Br/Ag(100) surface shows a very similar  
relaxation of the surface metal layers as that of the same configuration
for the Br/Au(100) surface. Both undergo an inward relaxation of the top
layer and slight outward relaxations of the second and third layers.
Similar relaxed structures are also found for the clean Ag(100) and
Au(100) surfaces (see Table~I).

Our calculations show that the second metal layer undergoes a small
buckling with the adsorption of Br in the $H_4$ configurations. The atoms
in the second layer that are immediately below the $H_4$ sites are 
observed to shift slightly up towards the surface, while the other atoms 
in that layer shift up by only on the order of 0.001 {\AA} and hence   
essentially keep their bulk positions. The spacing between these two
sub-layers is found to be 0.02 {\AA} and 0.04 {\AA} for the Br/Ag(100) and
Br/Au(100) surfaces. The distance between Br and the second-layer metal   
atom just below it is still far larger than the bond length between Br and
its nearest-neighbor metal atoms in the top layer. Thus a pseudo-five-fold
coordination, which has been observed in $c(2\times2)$ overlayer
structures on bcc metal surfaces, \cite{Griffiths,Bessent} does not exist
for the Br/Ag(100) and Br/Au(100) surfaces. This buckling may give rise to
an effective Br-Br interaction, mediated through the surface strain field.
The top metal layer in the $H_4$ configurations still shows a slight
inward relaxation, similar to the cases of the clean surfaces and the
$B_2$ configurations.

The top metal layer of the $T_1$ configuration of the Br/Au(100) surface
also shows a small buckling. The Au atoms in the top layer that are bonded
to Br are observed to undergo a larger inward relaxation than the other
half of the Au atoms in the top metal layer. The corresponding buckling
is, however, very large for the $T_1$ configuration of the Br/Ag(100)
surface. The distance between the two sublayers formed from the top Ag
layer is 0.75 {\AA}, indicating a zigzag surface reconstruction.

Finally, in Table~IV we show results of convergence checks for the  
total-energy differences. Such checks are particularly important for the
Br/Au(100) surface due to the small value of the total-energy difference
between the $H_4$ and $B_2$ configurations. Calculations with a higher
cutoff energy (30 Ry) obtained total-energy differences within 1 meV of
those from calculations with a cutoff energy of 20 Ry. Increasing the
number of special {\bf k} points from 36 to 56, increasing the slab
thickness from 7 to 9 metal layers, and increasing the vacuum region in  
the supercell from 5 to 7 layers, all changed the results by only a few
meV. Supercells with 5 metal layers are seen to cause errors in the 
total-energy differences of $\sim$30 meV for Br/Ag(100) and $\sim$10 meV  
for Br/Au(100). The use of 20 special {\bf k} points also causes an error 
of $\sim$10 meV. Therefore, it is necessary to employ supercells with at  
least 7 metal layers and 36 special {\bf k} points for obtaining the
total-energy differences with errors smaller than 10 meV. The distances
between Br and its nearest-neighbor metal atoms were also checked. We
found that the changes of these distances were smaller than 0.01 {\AA}
over the ranges of cut-off energies between 20 and 30 Ry, numbers of {\bf 
k} points between 20 and 56, and numbers of metal layers between 5 and 9
in the supercells, indicating that the bond lengths are not very sensitive
to the choice of computational parameters.

\subsubsection{Electronic properties and bonding character}

In order to better understand the differences between the bonding of
bromine on the Ag(100) and Au(100) surfaces, we calculated the total
electronic density of states 
(DOS), the DOS projected onto individual
atoms and specific atomic states, and the charge transfer between bromine
and the substrate.

Figure~2 shows the total DOS for the Br-adsorbed Ag(100) and Au(100)
surfaces. For comparison, the total DOS for the clean relaxed Ag(100) and
Au(100) surfaces are also shown. The peaks of the DOS curves for the clean
surfaces represent the main features of the $s$ and $d$ states of the
substrates and remain essentially at the same positions when bromine atoms
are adsorbed. New states are, however, found to be located at between   
$-$15 eV and $-$13 eV relative to the Fermi level in the DOS curves of the
Br-adsorbed Ag(100) and Au(100) surfaces. These states are predominantly
the bromine $3s$ states with small contributions from the $s$ and $d$    
states of the substrate, as seen from the curves for the DOS projected    
onto the specific atomic states of the adsorbate and the substrate (shown
in Fig.~3). Significant changes of the total DOS in the higher energy
range close to the Fermi level (above $-$2.5 eV and $-$1.5 eV for the    
Br/Ag(100) and the Br/Au(100) surfaces, respectively) are also observed

\vspace{1.cm}
\begin{figure}[thbp]
\epsfbox{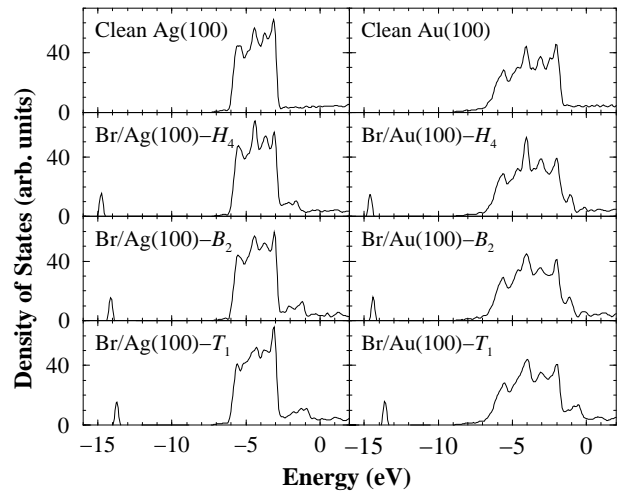}
\caption{Total density of states for the clean relaxed and the Br-adsorbed
Ag(100) and Au(100) surfaces. The Fermi level is at 0 eV. In this figure,
as well as in Figs. 3 and 4, the curves are obtained from calculations
with supercells containing a 7-layer slab and a 7-layer vacuum region, a
cutoff energy of 20 Ry, and 36 special {\bf k}-points.}
\label{autonum}
\end{figure}

\vspace{1.50cm}
\begin{figure}[thbp]
\epsfbox{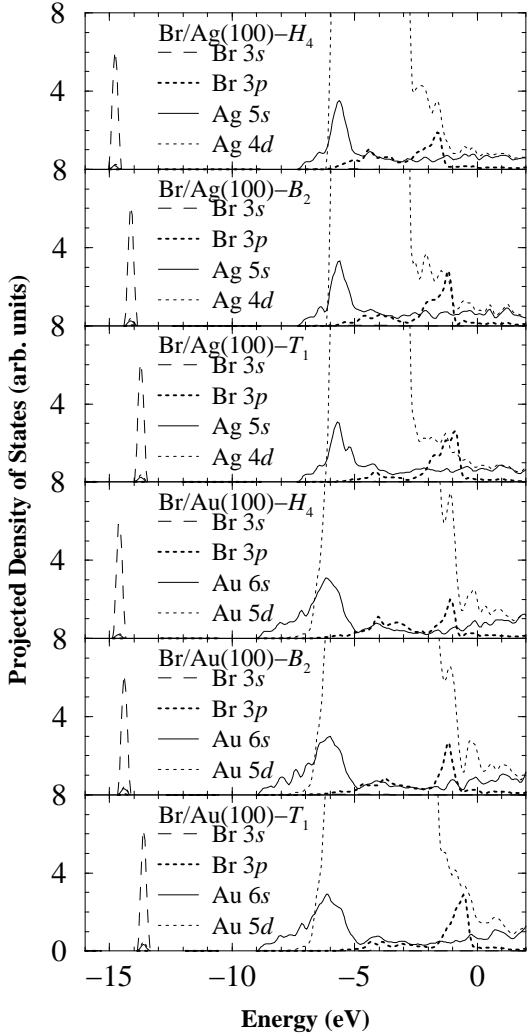}
\caption{The density of states projected onto the Br $3s$, Br $3p$, Ag
$5s$, Ag $4d$, Au $6s$, and Au $5d$ states for the $H_4$, $B_2$, and $T_1$
configurations of the Br-adsorbed Ag(100) and Au(100) surfaces.}
\label{autonum}
\end{figure}
\vspace{.5cm}

\parindent=0mm

when the DOS curves for the clean surfaces are compared with those for the
$H_4$, $B_2$, and $T_1$ configurations of the Br-adsorbed surfaces (see
Fig.~2). The electronic states in the higher energy range are composed
mostly of the bromine $3p$ states and the Ag 4$d$ (or Au 5$d$) states,
with some contributions coming from the Ag 5$s$ (or Au 6$s$) states (see
Fig.~3).

\parindent=4mm

In Fig.~4, we show the DOS projected onto bromine for the systems before
and after adsorption. The results for the systems before adsorption were
calculated by employing the supercell of a 7-layer slab and a 7-layer
vacuum region. A bromine layer with a c($2\times2$) periodicity was kept 
fixed in the middle of the vacuum region (located at $\sim$ 7 {\AA} above
the surface) so that there was essentially no interaction between Br and
the substrate. The peaks located in the lower and higher energy ranges in
the projected DOS before adsorption are due to the Br $3s$ and $3p$  
states, respectively. The slight broadening of the $3p$ states reflects
the weak $3p$-$3p$ interaction between neighboring bromine atoms. When   
bromine is adsorbed, both the $3s$ and $3p$ states shift down in energy  
due to the bonding between bromine and the substrate. A broadening of the
Br $3p$ states is also observed and can be attributed to the hybridization
of the bromine $3p$ states with the $s$- and $d$-bands of the substrates
(see Fig.~3). The Br $3s$ states are also seen to mix slightly with the 
$s$ and $d$ states of the substrate (see also Fig.~3). The hybridization
of the Br $3s$ and $3p$ states with the electronic states of the 
substrate suggests covalent bonding between bromine and the Ag(100) and
Au(100) surfaces.

The bonding of Br with the Ag(100) and Au(100) surfaces is also found to
be associated with a charge transfer from the substrate to Br. To obtain a
rough estimate of the charge transfer, we calculated the change of the   
charge for a bromine atom upon adsorption by integrating the difference of
the corresponding charge densities over a sphere with a radius of 1.28
{\AA} around the atom. \cite{RADIUS} We found that 0.15 and 0.14 electrons
were transfered from the Ag(100) and Au(100) surfaces, respectively, to
the bromine atom. The amount of the charge transfer was found to be
basically the same for the $H_4$, $B_2$, and $T_1$ configurations. These
results are consistent with the data for the DOS projected onto the Br
atom. By integrating the $3p$ contributions up to the Fermi level, we
observe that more $3p$ states are occupied in the Br-adsorbed surfaces
than in the systems before adsorption (see Fig.~4). A recent periodic GGA
calculation with a local basis set for the adsorption of chlorine on the 
Ag(111) surface also found that a slight charge ($\sim$ 0.2 electrons) was
transferred from the Ag(111) substrate to the chlorine atom. \cite{Doll}  
Experimental measurements of the electrosorption valency of Br adsorbed  
on Ag(100) report values of approximately $-$0.70 to $-$0.75,
\cite{Wandlowskinew,Kope1,Mitchell1,Mitchell2}
corresponding to a residual charge of 0.25 to 0.30 electrons on the
adsorbed
Br. These values are considerably larger than our calculated
charge of 0.15 electrons. The discrepancy may be due to the fact that our
calculations were performed for systems in vacuum. In an electrochemical
environment, the net charge associated with the adsorbate might be very
different from that in vacuum, due to solvation. However, the discrepancy
might also be attributed to inaccuracies in the theoretical and
experimental methods used to estimate the charge.

The difference in the bonding strength of the bromine with the substrate
between the different configurations directly affects their relative
stability. Based on our DOS data, we provide a qualitative explanation of
the difference in bonding strength between the $H_4$ and $B_2$
configurations. The Br $3s$ and $3p$ states in the Br/Ag(100)-$H_4$
configuration are significantly lower in energy than in the
Br/Ag(100)-$B_2$ configuration (see Fig.~4). In addition, the intensity of
the lower part of the $3p$ states (below approximately $-$3 eV) is larger
for Br/Ag(100)-$H_4$ than for Br/Ag(100)-$B_2$. Both facts suggest a
stronger bonding for the $H_4$ configuration on Ag(100). This is expected
since there are more direct bonding neighbors for the bromine atom at the
$H_4$ site. On the other hand, the Br 

\vspace{1.0cm}
\begin{figure}[thbp]
\epsfbox{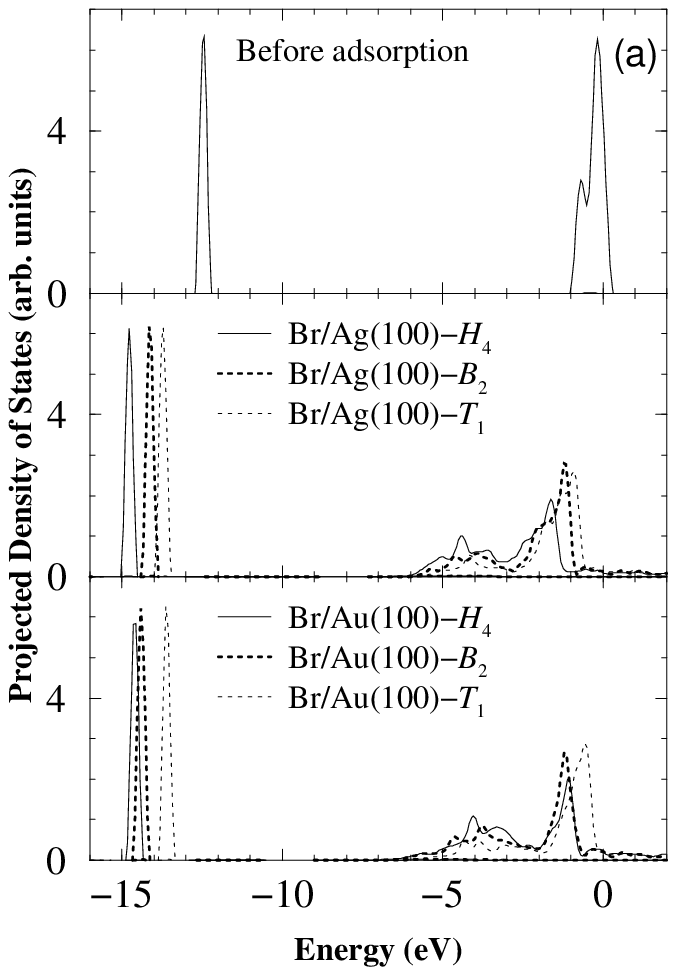}
\end{figure}

\vspace{-.50cm}

\begin{figure}[thbp]
\epsfbox{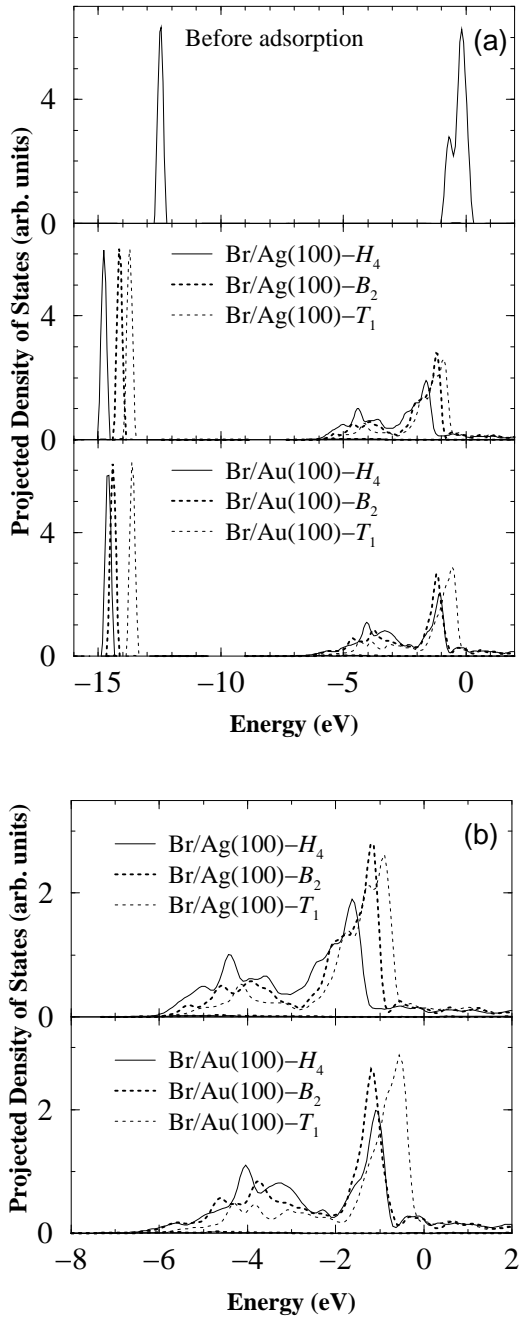}
\caption{The density of states projected onto Br for the $H_4$, $B_2$, and
$T_1$ configurations of the Br-adsorbed Ag(100) and Au(100) surfaces (a)
over a larger energy range ($-16$ eV to 2 eV) in which both the $3s$ and
$3p$ states are shown, and (b) over a smaller energy range ($-8$ eV to 2
eV) where only the $3p$ states are presented. Also shown in (a) is the DOS
projected onto Br without adsorption.}
\label{autonum}
\end{figure}

\parindent=0mm

$3s$ states in the Br/Au(100)-$H_4$ configuration are
only slightly lower in energy than in the
Br/Au(100)-$B_2$ configuration. While the Br $3p$ states extend over
almost the same range in energy for the Br/Au(100)-$H_4$ and -$B_2$
configurations, they have slightly larger intensity in the lower part
(below $\sim$ $-$2 eV) and smaller intensity in the higher part (above
$\sim$ $-$2 eV) for the Br/Au(100)-$H_4$ configuration than for the
Br/Au(100)-$B_2$ configuration. While the $H_4$ configuration thus has a
stronger covalent bonding for both the Br-adsorbed Ag(100) and Au(100)
surfaces, the difference in bonding strength between the $H_4$ and the
corresponding $B_2$ configurations is smaller for the Au(100) surface than
for the Ag(100) surface. This is probably due to the fact that the Au $6s$
and $5d$ electrons are more delocalized than the Ag $5s$ and $4d$
electrons, and the bonding strength is expected to be less sensitive to
the bonding sites for substrates with more delocalized electrons. In
addition to the stronger covalent bonding, the Coulomb attraction
resulting from the charge transfer in the $H_4$ configuration is also
stronger than the corresponding $B_2$ configuration for both the
Br/Ag(100) and Br/Au(100) systems, due to the shorter distance between
bromine and the surface in the $H_4$ configuration (see Table~III).

\parindent=4mm

It is clear that there is a delicate competition between the attractive
and repulsive interactions in each configuration. In particular, the
core-core repulsion between bromine and the substrate, which is irrelevant
to the electronic DOS but makes a contribution to the total energy of the
system, in the $H_4$ configuration is stronger than in the $B_2$
configuration. The core-core repulsive energy is calculated as an Ewald  
sum \cite{Payne,Ihm} (see the third column of Table~V). The total energy,
as determined in our DFT calculations, contains as separate parts the
electronic and the core-core Coulomb contributions. Differences of the
electronic and the core-core contributions to the total energy between the
$H_4$ and the $B_2$ configuration are presented in Table~V. We note that
consideration of the electronic contributions alone does not properly
address the opposite order of the total-energy difference (ie, the
binding-energy difference) between the $H_4$ and $B_2$ configurations for
the the Br/Ag(100) and Br/Au(100) surfaces. The core-core interactions
need to be included. For both the Br/Ag(100) and Br/Au(100) surfaces, the
electronic contribution favors the $H_4$ configuration, while the
core-core contribution favors the $B_2$ configuration. In the Ag(100)
case, the core-core energy, which is higher for $H_4$ than for $B_2$, is
more than compensated by the lower electronic energy for the $H_4$
configuration, resulting in $H_4$ being the preferred bonding site for Br
on Ag(100). For the Br/Au(100) system, however, the lower electronic
energy for the $H_4$ configuration only partially compensates the higher
core-core energy for Br/Au(100)-$H_4$. As a result, the $B_2$
configuration is lower in total energy than the $H_4$ configuration for
the Br-adsorbed Au(100) surface. The small magnitudes of the total-energy
differences, compared to the individual electronic and core-core
contributions, strongly emphasize the need for very accurate energy
calculations and careful convergence checks.

\section{Conclusions}

The theoretical approach of supercell models combined with
first-principles total-energy DFT pseudopotential methods has reproduced 
experimental measurements of preferred adsorption sites for Br-chemisorbed
Ag(100) and Au(100) surfaces.

We have shown that while the hollow-site configuration is more stable on
the Br/Ag(100) surface (by 210 meV/adatom over the bridge-site structure),
the bridge-site configuration is more stable than the corresponding 
hollow-site structure by 60 meV/adatom on the Br/Au(100) surface. The
calculations also predict that the one-fold on-top configuration is the  
least stable structure on both surfaces (560 meV and 300 meV higher than
the corresponding most stable structure for the Br/Ag(100) and Br/Au(100)
surfaces, respectively). Other aspects of the geometries of the Br/Ag(100)
and Br/Au(100) systems have also been determined and are shown to be in   
excellent agreement with the available experimental data.

The bond between Br and the substrate is found to be covalent with a
slight polarization due to a small charge transfer from the substrate to
the bromine. The chemical bonding between Br and the substrate is shown to
be stronger in the $H_4$ configuration than in the $B_2$ configuration.
Compared with the Br/Ag(100) surface, however, the Br/Au(100) surface
exhibits a reduced difference in the bonding strength between the $H_4$   
and $B_2$ configurations. The core-core Coulomb interaction is found to be
higher for the $H_4$ configuration than for the $B_2$ configuration. The
detailed balance between the electronic and the core-core contributions to
the total energy determines $H_4$ and $B_2$ as the the preferred bonding
site on the Ag(100) and Au(100) surfaces, respectively.

Our work demonstrates that the use of extended surface models and careful
convergence checks are critical for obtaining reliable information on the
Br/Ag(100) and Br/Au(100) systems from {\it ab initio} calculations.

\acknowledgments

We thank S.~J.~Mitchell and L.~G. Wang for helpful discussions. We also
thank S.~P.~Lewis, M.~A. Novotny and G. Brown for comments on the
manuscript. This work was supported by the National Science Foundation
under grant No. DMR-9981815 and by Florida State University through the
Center for Materials Research and Technology and the School of
Computational Science and Technology.

\onecolumn

\begin{table}
\widetext
\caption{Relaxation of the $H_4$, $B_2$, and $T_1$ configurations of the
Br/Ag(100)-$c(2\times2)$ and Br/Au(100)-$c(2\times2)$ surfaces. Also shown
are the vertical distance ($d_z$, in units of \AA) between Br and the
plane of the centers of the top-layer atoms on the surface, and the
distance ($d$, in units of \AA) between Br and its nearest-neighbor metal
atom(s). The change in spacing between layers $i$ and $j$ is denoted by
$\Delta d_{ij}$. When sub-layers are present, the changes are denoted by
$\Delta d_{ij}$ and $\Delta d_{ij}'$, with $\Delta d_{ij}$ being the
larger in magnitude. $d_{0}$ is the same as in Table~I. The computational
details are the same as in Table~II.}
\begin{tabular}{lccccccc}
&$d_z$&$d$&${\Delta d_{12}}/d_0$ (\%)&${\Delta d_{12}'}/d_0$ (\%)&${\Delta
d_{23}}/d_0$ (\%)&$%
{\Delta d_{23}'}/d_0$ (\%)&${\Delta d_{34}}/d_0$ (\%)\\
\tableline
Br/Ag(100)-$H_4$&1.91&2.82&$-$0.6&0.1&0.8&0.1&$-$0.0\\
Br/Ag(100)-$B_2$&2.16&2.61&$-$0.9&&0.5&&0.2\\
Br/Ag(100)-$T_1$&2.48&2.48&10.4&$-$7.7&0.2&&$-$0.1\\
\tableline
Br/Au(100)-$H_4$&2.01&2.89&$-$1.6&0.2&1.4&$-$0.3&0.4\\
Br/Au(100)-$B_2$&2.18&2.62&$-$0.9&&0.4&&0.3\\
Br/Au(100)-$T_1$&2.46&2.46&$-$1.2&$-$0.5&0.5&&0.4\\
\end{tabular}
\end{table}   

\begin{table}
\mediumtext
\caption{Convergence checks for the total energy differences (in units of
eV per unit cell) between different configurations with respect to the
cutoff energy ($E_{cut}$, in units of Ry), the number of metal layers
($N_m$) in the supercell, the thickness of the vacuum region ($N_v$, in   
units of number of bulk metal layers), and the number of {\bf k} points in
the surface Brillouin zone ($N_k$).}
\begin{tabular}{cccccccc}
$E_{cut}$&$N_m$&$N_v$&$N_k$&$E_{H_4}-E_{B_2}$&$E_{H_4}-E_{T_1}$&$E_{H_4}-%
E_{B_2}$&$E_{B_2}-E_{T_1}$\\
&&&&Br/Ag(100)&Br/Ag(100)&Br/Au(100)&Br/Au(100)\\
\tableline
20&5&7&20&$-$0.191&$-$0.526&$+$0.061&$-$0.282\\
20&7&7&20&$-$0.222&$-$0.556&$+$0.048&$-$0.285\\
20&7&7&36&$-$0.211&$-$0.552&$+$0.056&$-$0.291\\
30&7&7&36&$-$0.212&$-$0.552&$+$0.057&$-$0.291\\
20&7&5&36&$-$0.210&$-$0.550&$+$0.057&$-$0.289\\
20&7&7&56&$-$0.210&$-$0.551&$+$0.059&$-$0.294\\
20&9&7&56&$-$0.213&$-$0.557&$+$0.058&$-$0.302\\
\end{tabular}
\end{table}

\begin{table}
\narrowtext
\caption{Differences (in eV per unit cell) of two energy contributions
to the total energy between the $H_4$ and $B_2$ configurations. $E^e$ and
$E^{cc}$ are the electronic and core-core Coulomb contributions,
respectively. Also shown is the total-energy difference
($E_{H_4}^{tot}-E_{B_2}^{tot}$). The results are obtained from
calculations with supercells containing a 7-layer slab and a 7-layer
vacuum region, a cutoff energy of 20 Ry, and 36 special {\bf k}-points.}
\begin{tabular}{lcccc}
&$E_{H_4}^{e}-E_{B_2}^{e}$&$E_{H_4}^{cc}-E_{B_2}^{cc}$
&$E_{H_4}^{tot}-E_{B_2}^{tot}$ \\
\tableline
Br/Ag(100)&$-$429.230&$+$429.018&$-$0.212\\
Br/Au(100)&$-$371.156&$+$371.213&$+$0.057\\
\end{tabular}
\end{table}


\begin{references}
\bibitem{RADIUS}
The value of 1.28 {\AA} was chosen to be intermediate between the atomic
(1.14 {\AA}) and ionic (1.95 {\AA}) radii of bromine, but closer to the
atomic radius.
\end{references}
\end{document}